\documentclass{PoS}

\PoS{PoS(LAT2005)069}
\usepackage{amsmath}

\renewcommand{\S}{{\mathrm{S}}}
\newcommand{\V}{{\mathrm{V}}}
\newcommand{\A}{{\mathrm{A}}}
\renewcommand{\P}{{\mathrm{P}}}
\newcommand{\T}{{\mathrm{T}}}

\newcommand{\ii}{{\mathrm{i}}}

\newcommand{\C}{{\mathrm{C}}}

\newcommand{\bfV}{\mathbf{V}}
\newcommand{\bfA}{\mathbf{A}}

\title{The Quark Structure of Pentaquarks }

\ShortTitle{The Quark Structure of Pentaquarks }

\author{O.~Jahn, \speaker{J.W.~Negele},  and D.~Sigaev\\
  Center for Theoretical Physics, Massachusetts Institute of Technology, Cambridge, MA 02139, USA \\
        E-mail: 
        \email{jahn@mit.edu},
        \email{negele@mitlns.mit.edu},
        \email{sigaev@mit.edu}}

\abstract{Motivated by the possible observation of the $\Theta^+(1530)$, we study the quark structure of pentaquark states in quenched lattice QCD.  The complete set of 19 local sources  that have the proper symmetry for positive or negative parity isoscalar pentaquarks is constructed, as well as a nonlocal source composed of two displaced ``good'' diquarks.  Quantitative structure information is determined from diagonalizing the 19-dimensional correlation matrix and from calculating the overlaps of sources with the lattice eigenstates. The volume dependence of the overlap is studied to differentiate between scattering and localized resonant states.  The positive parity state has a small component of two ``good'' diquarks, and its energy is too much higher than the negative parity state to be a candidate for the   $\Theta^+(1530)$.}

\FullConference{XXIIIrd International Symposium on Lattice Field Theory\\
		 25-30 July 2005\\
		 Trinity College, Dublin, Ireland}

\begin{document}

\section{Introduction}

Because of the strong spin and color dependent interaction between quarks, theorists have long sought exotic bound states whose dominant quark content differs from the familiar quark-antiquark mesons and three quark baryons.  The recent apparent observation of the  $\Theta^+$ pentaquark with minimal quark content $uudd\bar s$ focussed renewed interest on such exotics. Even though the experimental evidence is now  suspect, as summarized in a recent review\cite{Danilov:2005kt}, it is interesting to explore these states theoretically.  

Lattice field theory  is the only quantitative alternative to models such as the chiral soliton model\cite{Diakonov:1997mm} that motivated the search for the $\Theta^+$ and the physically appealing diquark picture\cite{Jaffe:2003sg}, and a number of lattice calculations have now been published\cite{
Csikor:2003ng,
Sasaki:2003gi,
Chiu:2004gg,
Mathur:2004jr,
Ishii:2004qe,
Alexandrou:2004ws,
Takahashi:2004sc,
Lasscock:2005tt,
Csikor:2005xb,
Alexandrou:2005gc,
Takahashi:2005uk}.
With one exception, all these works have considered at most three simple sources for spin-1/2 pentaquarks, the diquark source
\begin{equation*}
 \Pi^{\textrm{Diquark}} =  \epsilon_{abc}\epsilon_{bef}\epsilon_{cgh}(u^{Te}Cd^f)(u^{Tg}C \gamma_5d^h)C\bar s^{Tc}
\end{equation*}
the K-N source 
\begin{equation*}
 \Pi^{\textrm{KN}} = \epsilon_{abc}(u^{Ta}C\gamma_5d^b) \gamma_5 u^c(\bar s^e \gamma_5 d^e),
\end{equation*} 
and the color-fused K-N source
\begin{equation*}
 \Pi^{\textrm{cfKN}} = \epsilon_{abc}(u^{Ta}C\gamma_5d^b) \gamma_5 u^e(\bar s^e \gamma_5 d^c).
\end{equation*} 
Reference\cite{Csikor:2005xb} also considers a spatially displaced K-N source and a spatially displaced diquark source of the form discussed below.  All these calculations had the property that one could not calculate three eigenstates for each parity and investigate the volume dependence of these eigenstates.  Hence, the objective of the present work is to construct all the local sources of each parity, calculate the lowest few eigenstates of the correlation matrix constructed from them, and examine the masses, expansion coefficents, volume dependence, and overlaps with the physical states of these eigenstates.

\section{Sources}

The 19 independent local sources are constructed\cite{Jahn:2005} by first coupling 4 quarks to a color triplet and isosinglet and then coupling the result with an $\bar s$ to form a spin-1/2 color singlet.This yields 19 rotationally covariant operators on each time slice. These sources may be written succinctly in terms of the pseudoscalar, scalar, vector, axial vector and tensor diquark operators
\begin{align*}
  Q^{\P}_{c} &= \epsilon_{c a b} \, \epsilon^{i j} \, (q^a_i C q^b_j) 
 & Q^{\S}_{c} &= \epsilon_{c a b} \, \epsilon^{i j} \, (q^a_i C\gamma_5 q^b_j) \\
  Q^{\V}_{c\,\mu} &= \epsilon_{c a b} \, \epsilon^{i j} \, (q^a_i C\gamma_5\gamma_\mu q^b_j) 
 & Q^{\A}_{c\,n\,\mu} &= \epsilon_{c a b} \, (\tau_2\tau_n)^{i j} \, (q^a_i C\gamma_\mu q^b_j) \\
  Q^{\T}_{c\,n\,\mu\nu} &= \epsilon_{c a b} \, (\tau_2\tau_n)^{i j} \, (q^a_i C\sigma_{\mu\nu} q^b_j) \;, &  &
\end{align*}
and using the notation  $s^C = C \bar s ^T$ and
\begin{equation*}
  Q \wedge Q' \cdot s^\C \equiv 
  \begin{cases}
    \epsilon^{e f g} \, Q_f \, Q'_g \, s^\C_e &\quad\text{for isoscalar $Q$, $Q'$} \\
    \epsilon^{e f g} \, Q_{f n} \, Q'_{g n} \, s^\C_e &\quad\text{for isovector $Q$, $Q'$.}
  \end{cases}
\end{equation*}
This construction yields the following seven operators containing isoscalar diquarks
\\[-.6cm]
\begin{align*}
  \Pi^{\P\S}_{\P} &= \phantom{-\epsilon_{i j k}} \; 
  Q^{\P} \wedge Q^{\S} \cdot s^\C 
 &
  \Pi^{\S\S'}_{\S} &= \phantom{-\epsilon_{i j k}} \; 
  Q^{\S} \wedge Q^{\V}_{4} \cdot \gamma_5\gamma_4 s^\C
\\
  \Pi^{\S\bfV}_{\bfV} &= \phantom{-\epsilon_{i j k}} \; 
  Q^{\S} \wedge Q^{\V}_{i} \cdot \gamma_5\gamma_i s^\C
&
  \Pi^{\P\S'}_{\P} &= \phantom{-\epsilon_{i j k}} \; 
  Q^{\P} \wedge Q^{\V}_{4} \cdot \gamma_4 s^\C
\\
  \Pi^{\P\bfV}_{\bfA} &= \phantom{-\epsilon_{i j k}} \; 
  Q^{\P} \wedge Q^{\V}_{i} \cdot \gamma_i s^\C
&
  \Pi^{\S'\bfV}_{\bfV} &= -\epsilon_{i j k} \;
  Q^{\V}_{4} \wedge Q^{\V}_{i} \cdot 
  \sigma_{j k} s^\C
\\
  \Pi^{\bfV\bfV}_{\bfA} &= \phantom{-}\epsilon_{i j k} \;
  Q^{\V}_{i} \wedge Q^{\V}_{j} \cdot 
  \sigma_{k 4} s^\C  & & 
\end{align*}
and the following twelve operators containing
 isovector diquarks
\begin{align*}
  \Pi^{\P'\bfA'}_{\bfV} &= -\phantom{\epsilon_{i j k}} \; 
  Q^{\A}_{4} \wedge Q^{\T}_{i 4} \cdot \gamma_5\gamma_i s^\C
&
  \Pi^{\bfA\bfA'}_{\S} &= \phantom{-\epsilon_{i j k}} \; 
  Q^{\A}_{i} \wedge Q^{\T}_{i 4} \cdot \gamma_5\gamma_4 s^\C
\\
  \Pi^{\bfA\bfV'}_{\bfV} &= \phantom{-\epsilon_{i j k}} \; 
  Q^{\A}_{i} \wedge Q^{\T}_{i j} \cdot \gamma_5\gamma_j s^\C
&
  \Pi^{\P'\bfV'}_{\bfA} &= -\epsilon_{i j k} \; 
  Q^{\A}_{4} \wedge Q^{\T}_{i j} \cdot \gamma_k s^\C
\\
  \Pi^{\bfA\bfV'}_{\P} &= \phantom{-}\epsilon_{i j k} \; 
  Q^{\A}_{i} \wedge Q^{\T}_{j k} \cdot \gamma_4 s^\C
&
  \Pi^{\bfA\bfA'}_{\bfA} &= -\epsilon_{i j k} \;
  Q^{\A}_{i} \wedge Q^{\T}_{j 4} \cdot \gamma_k s^\C
\\
  \Pi^{\P'\bfA}_{\bfA} &= -\epsilon_{i j k} \;
  Q^{\A}_{4} \wedge Q^{\A}_{i} \cdot 
  \sigma_{j k} s^\C
&
  \Pi^{\bfA\bfA}_{\bfA} &= -\epsilon_{i j k} \;
  Q^{\A}_{i} \wedge Q^{\A}_{j} \cdot 
  \sigma_{4 k} s^\C
\\
  \Pi^{\bfA'\bfA'}_{\bfA} &= \phantom{-}\epsilon_{i j k} \;
  Q^{\T}_{i 4} \wedge Q^{\T}_{j 4} \cdot 
  \sigma_{k 4} s^\C
&
  \Pi^{\bfA'\bfV'}_{\bfV} &= \phantom{-} \epsilon_{i j k} \;
  Q^{\T}_{l 4} \wedge Q^{\T}_{i l} \cdot 
  \sigma_{j k} s^\C
\\
  \Pi^{\bfV'\bfV'}_{\bfA} &= -\epsilon_{i j k} \;
  Q^{\T}_{i l} \wedge Q^{\T}_{j l} \cdot 
  \sigma_{4 k} s^\C
&
  \Pi^{\bfA'\bfV'}_{\P} &= \phantom{-}\epsilon_{i j k} \;
  Q^{\T}_{i 4} \wedge Q^{\T}_{j k} \cdot 
  s^\C.
\label{eq:Pirot19}
\end{align*}
In this notation, the diquark source is $  \Pi^{\textrm{Diquark}} = \Pi^{\P\S}_{\P}$ and the K-N and color fused K-N sources are
\begin{align*}
  \Pi^{\textrm{KN}} &= \frac{1}{16} \left(
  -\Pi^{\P\S}_{\P} 
  +\Pi^{\S\S'}_{\S}  
   +  \Pi^{\S\bfV}_{\bfV}
  + \frac12 \Pi^{\P\S'}_{\P} 
   + \frac12  \Pi^{\P\bfV}_{\bfA}
  + \frac\ii4 \Pi^{\S'\bfV}_{\bfV} 
   + \frac\ii4  \Pi^{\bfV\bfV}_{\bfA} 
  - \frac\ii4 \Pi^{\P'\bfV'}_{\bfA}  \right.  \\
   & \quad \quad \quad
 \left. - \frac\ii4 \Pi^{\bfA\bfV'}_{\P}
  - \frac\ii4 \Pi^{\bfA\bfA'}_{\bfA}
  + \frac\ii4 \Pi^{\P'\bfA}_{\bfA} 
  + \frac\ii4 \Pi^{\bfA\bfA}_{\bfA} 
  - \frac\ii4 \Pi^{\bfA'\bfA'}_{\bfA} 
  - \frac\ii4 \Pi^{\bfA'\bfV'}_{\bfV}
  - \frac\ii4 \Pi^{\bfV'\bfV'}_{\bfA}
  \right)
\\
  \Pi^{\textrm{cfKN}} &= \frac{1}{16} \left(
  +\Pi^{\P\S}_{\P} 
  -\Pi^{\S\S'}_{\S}  
   -  \Pi^{\S\bfV}_{\bfV}
  + \frac12 \Pi^{\P\S'}_{\P} 
   + \frac12  \Pi^{\P\bfV}_{\bfA}
  + \frac\ii4 \Pi^{\S'\bfV}_{\bfV} 
   + \frac\ii4  \Pi^{\bfV\bfV}_{\bfA} 
  - \frac\ii4 \Pi^{\P'\bfV'}_{\bfA}  \right.  \\
   & \quad \quad \quad
 \left. - \frac\ii4 \Pi^{\bfA\bfV'}_{\P}
  - \frac\ii4 \Pi^{\bfA\bfA'}_{\bfA}
  + \frac\ii4 \Pi^{\P'\bfA}_{\bfA} 
  + \frac\ii4 \Pi^{\bfA\bfA}_{\bfA} 
  - \frac\ii4 \Pi^{\bfA'\bfA'}_{\bfA} 
  - \frac\ii4 \Pi^{\bfA'\bfV'}_{\bfV}
  - \frac\ii4 \Pi^{\bfV'\bfV'}_{\bfA}
  \right)
\end{align*}
We also consider the nonlocal displaced diquark source   $Q^{\S}(x^-) \wedge Q^{\S}(x^+) \cdot \left(s^\C(x^+)  + s^\C(x^-) \right).   $

\section{Preliminary Lattice Results}
We construct the $19 \times 19$ correlation matrix 
$C_{\alpha \beta}(t) = \langle \Pi_\alpha(t) \bar \Pi_\beta(0) \rangle$ 
and solve the generalized eigenvalue equation\cite{Michael:1985ne,Luscher:1990ck}
$C(t)u = \lambda C(t_0) u$.  
The eigenvalue $ \lambda = e^{- m_{eff} (t-t_0)  }$ yields an effective mass that approaches the mass of the state at large $t$, and the eigenvector $u$ yields the expansion coefficients for the optimal variational wave function is the space generated by the sources $ |\psi\rangle = \sum_\alpha u_\alpha e^{-\frac{t_0}{2} H}\bar \Pi_\alpha | \Omega \rangle$

\begin{figure}[t]
\begin{center}
\includegraphics[width=.55\textwidth]{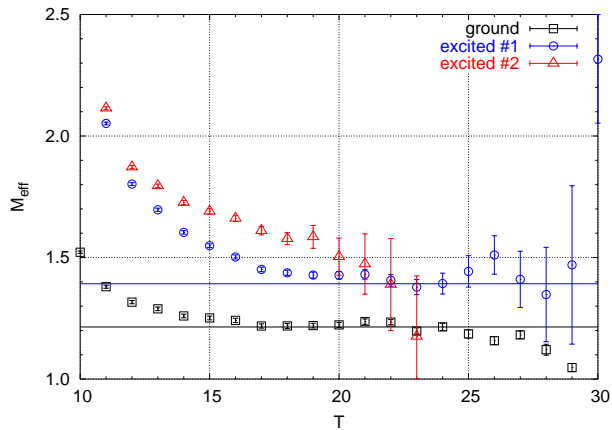}
\caption{Effective masses for the lowest three negative parity eigenstates of the $19 \times 19$ correlation matrix on the $16^3$ lattice }
\label{fig:mass}
\end{center}
\end{figure}
%
\begin{figure}[t]
\begin{center}
\includegraphics[width=.45\textwidth]{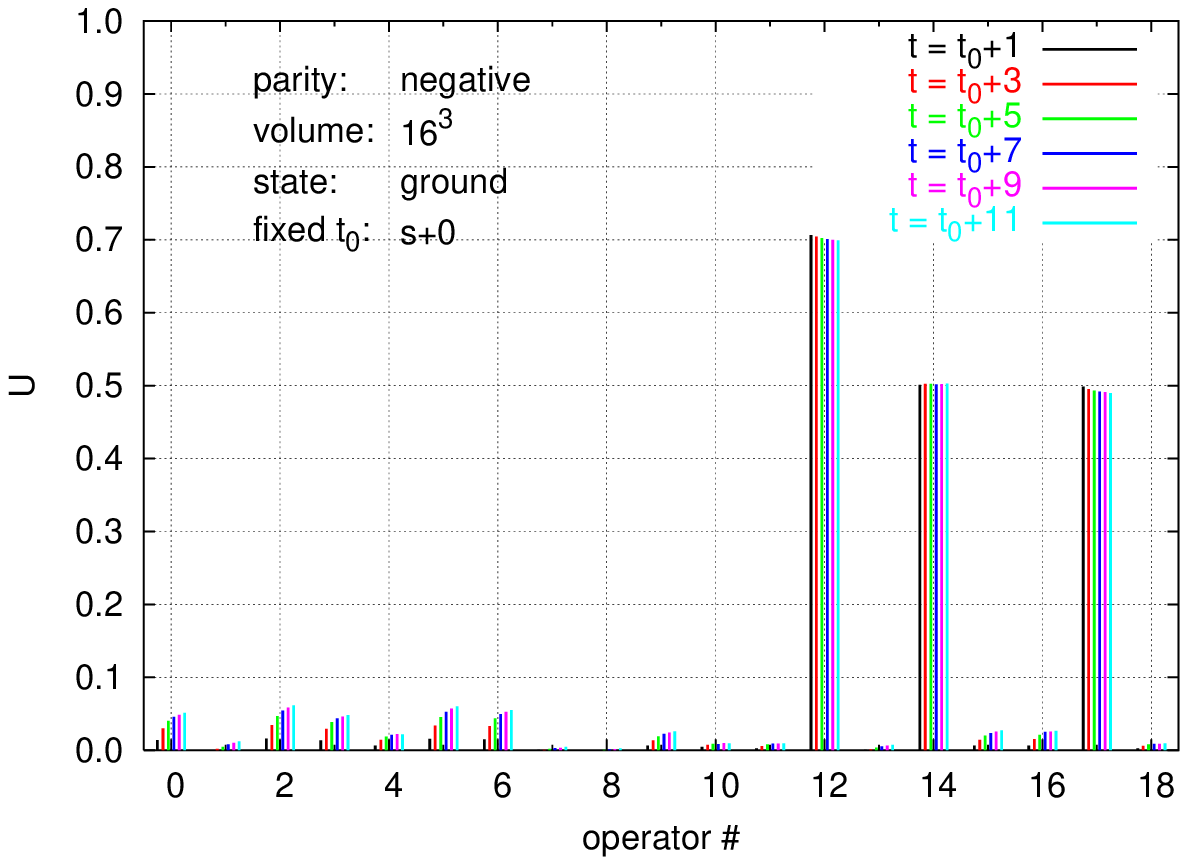}
\includegraphics[width=.45\textwidth]{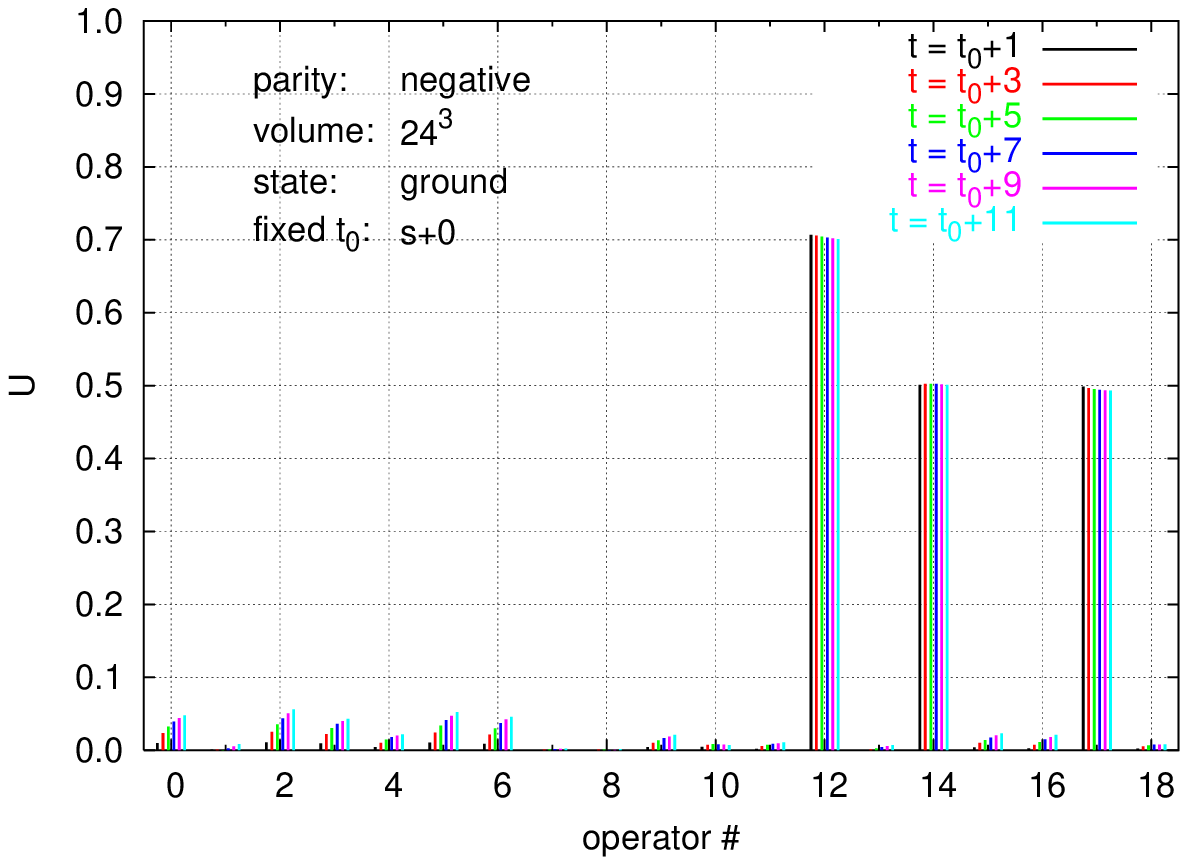}\\[-.1cm]
\includegraphics[width=.45\textwidth]{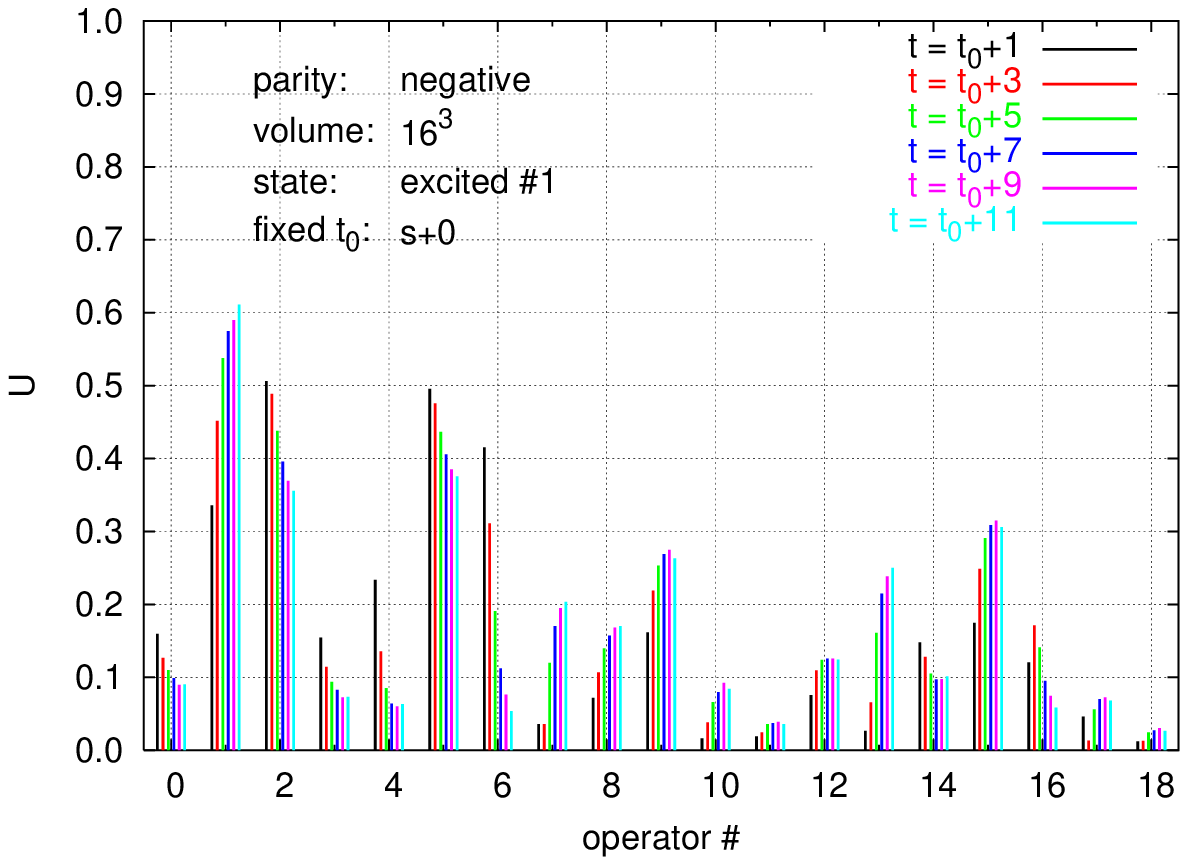}
\includegraphics[width=.45\textwidth]{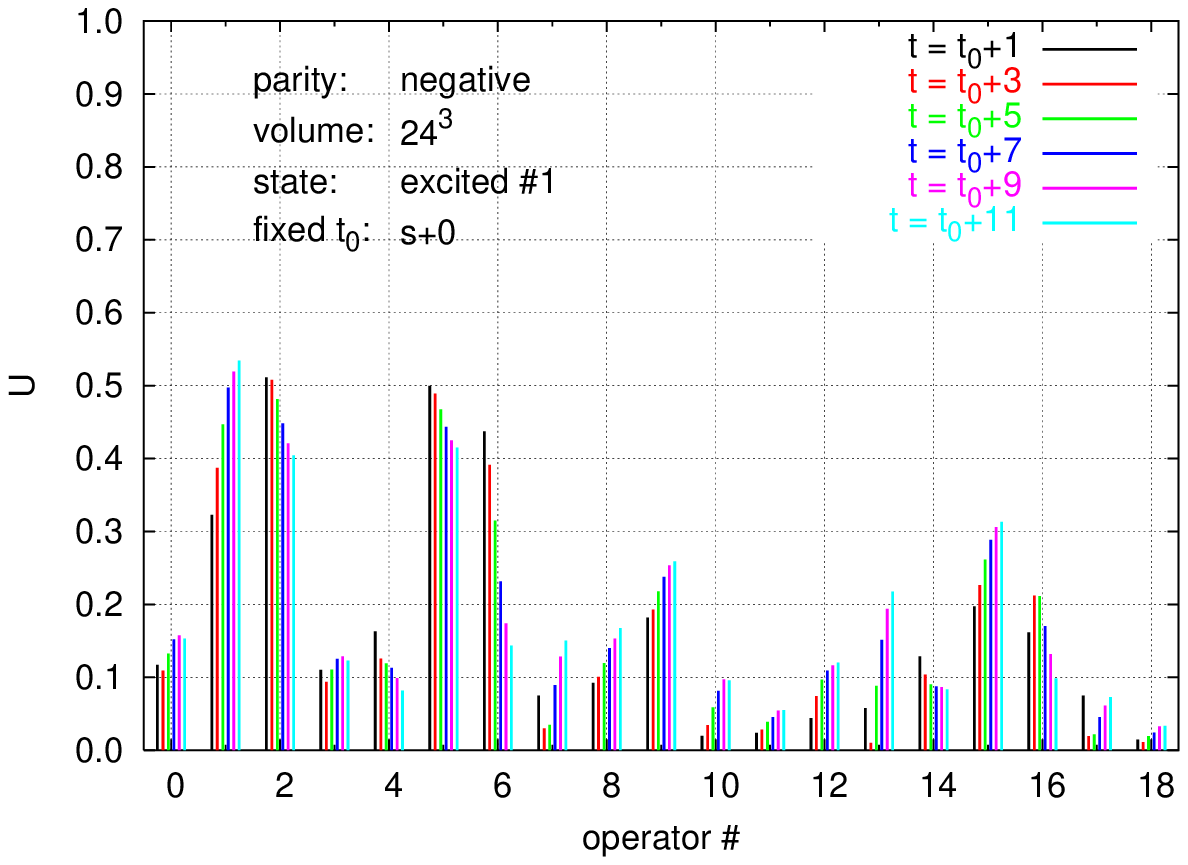}\\[-.1cm]
\includegraphics[width=.45\textwidth]{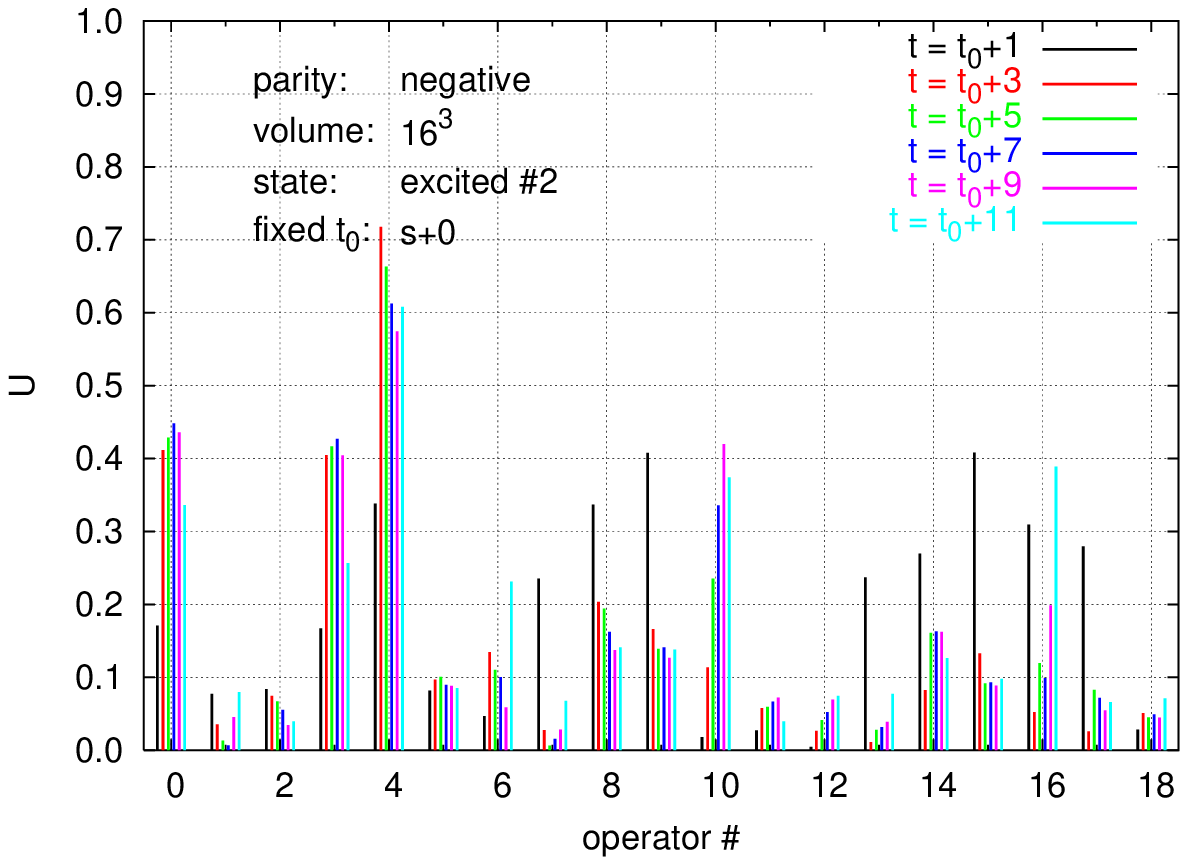}
\includegraphics[width=.45\textwidth]{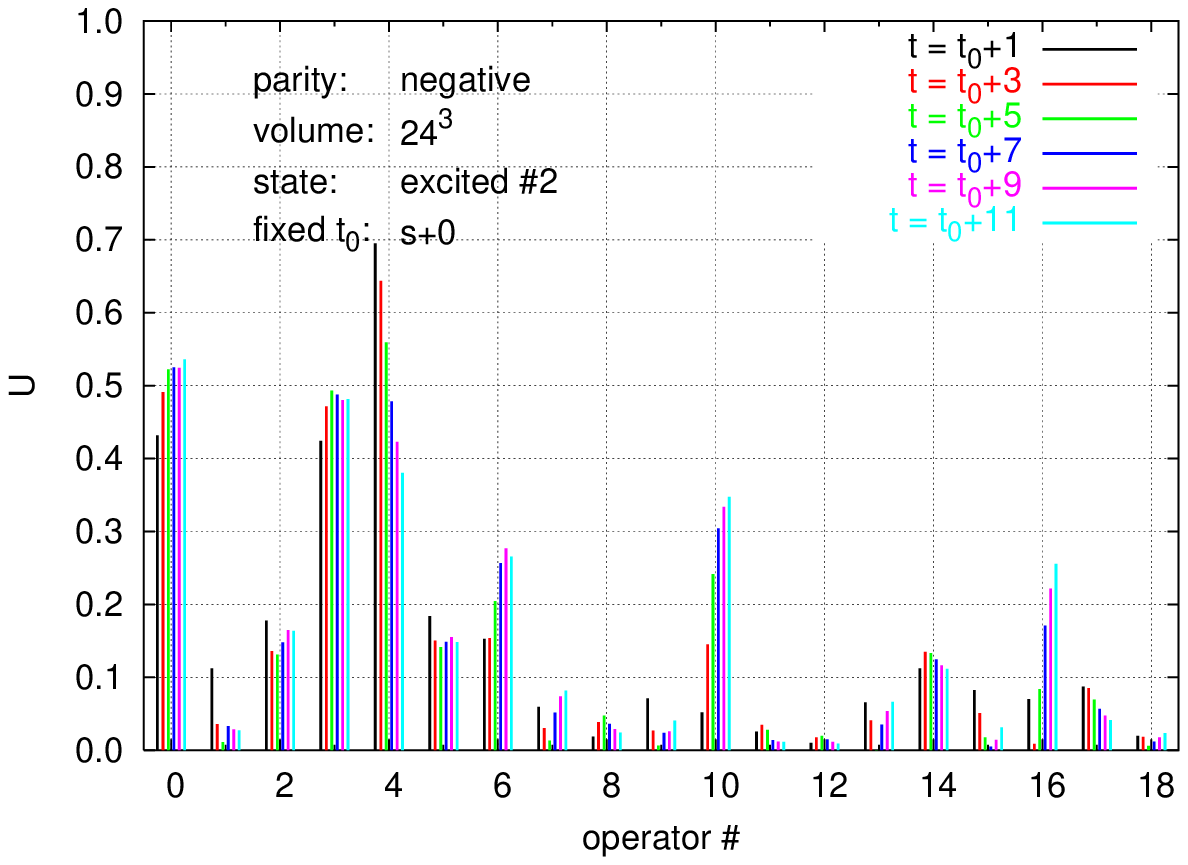}
\caption{Amplitude $u_\alpha$ of the state created by operator $\alpha$ in the lowest three eigenstates of the correlation matrix with $t_0$ located at the source.  The rows correspond to the ground and first two excited states. The left and right columns correspond to $16^3$ and $24^3$ respectively. The six bars for each operator denote the amplitude determined for $t$ ranging from $t_0 + 1 $  to $t_0 + 6$ .  }
\label{fig:amplitudes}
\end{center}
\end{figure}
%
\begin{figure}[t]
\begin{center}
\includegraphics[width=.55\textwidth]{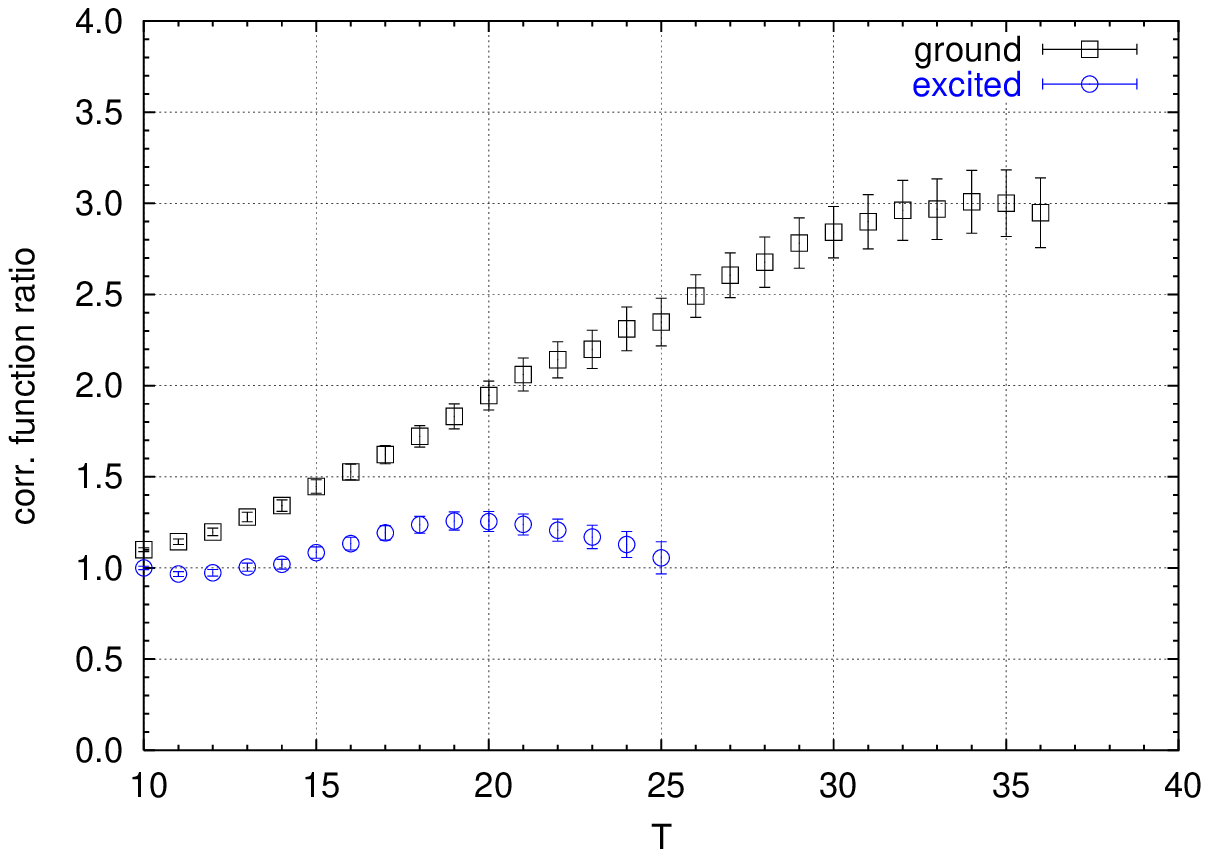}
\caption{Correlation function ratio $C_{16}(t)/C_{24}(t)$ for the negative parity ground state, denoted by squares, and first excited state, denoted by circles. The ground state approaches the ratio of the volumes, 3.375, indicative of a scattering state and the excited state ratio is approximately constant, consistent with a localized resonance. }
\label{fig:volume}
\end{center}
\end{figure}

Calculations were performed with quenched Wilson fermions for $\beta = 6$ and $m_\pi = 900$ MeV using Wuppertal smeared sources and sinks with $N=50$ and $\alpha = 3$ and using APE smeared links in the smeared sources. To obtain adequate statistics, we used 2688 configurations on a $16^3 \times 32$ lattice and 834 configurations on a $24^3 \times 32$ lattice.

Figure~\ref{fig:mass} shows the masses for the lowest three negative parity states, and it was not possible to separate the fourth state from the third state with present statistics. In the positive parity channel, the two lowest states were nearly degenerate and could not be distinguished statistically. 

Figure~\ref{fig:amplitudes}  shows the amplitudes, $u_\alpha$, for each of the 19 components in the optimal variational wave function for the two volumes.  There is only one local operator that has non-vanishing upper components for all five quarks and thus has a nonrelativistic limit, and this operator contributes (along with other terms) to three of the basis states.  The ground state is dominated by just these three states, and the linear combination is such that the nonrelativistic component is 98\% of the wave function, strongly suggesting that this state corresponds to a K-N scattering state.  The next two states are much more complicated, but have none of these K-N scattering components.  Note that the compositions of three states change negligibly with volume. 

The overlap between a localized state created by our local sources and a scattering state falls off with volume $V$ like $V^{-\frac{1}{2}}$, so the ratio of correlation functions at two volumes should vary as the inverse of the ratio of the volumes at large distance for scattering states, but remain unity for localized resonance states.  Figure~\ref{fig:volume} shows that the ground state, which we have already identified with the K-N scattering state, indeed does have a correlation function ratio that approaches the ratio of the volumes at large distances.  Interestingly, however, the first excited state has a ratio consistent with unity, suggestive of a possible localized state.  

Finally, it is useful to consider the overlaps between the trial functions and the physical eigenstates obtained by comparing the contribution of the ground state to the correlation function  and the full correlation function at   $t=0$. For reference, the overlaps for the pion on $16^3$ and $24^3$ lattices are 0.614(2) and 0.619(2) respectively - large and volume independent. For the nucleon, the corresponding  numbers are 0.605(4) and 0.608(3), again large and volume independent. For the negative parity ground state, the overlaps are 0.44(3) and 0.17(1), yielding a ratio 2.6(3) qualitatively consistent with the volume dependence expected for scattering.  For the first excited state, the overlaps are 0.17(1) and 0.12(1), yielding a ratio of 1.4(2) roughly consistent with a constant.  In contrast to the nucleon and pion however, the roughly 15\% overlap is very low, suggesting  far more complicated admixtures in the wave function.  The overlap of the displaced diquark source and the positive parity ground state is 0.03, indicating that this state is not dominated by a p-wave configuration of two good (scalar) diquarks.

These results indicate that diagonalization of the correlation matrix in a sufficiently large basis of pentaquark operators enables the study of several physical states, and that the combination of wave function amplitudes, overlaps, and volume dependence provide useful insight into the structure of each of these states.

\section {Acknowledgements}
This work was supported by the DOE Office of Nuclear Physics under contract DE-FC02-94ER40818.  Computations were performed on clusters at Jefferson Laboratory  using time awarded under the SciDAC initiative and on the MIT alpha cluster.

\end{document}